\documentclass[lettersize,journal]{IEEEtran}
\DeclareUnicodeCharacter{202F}{\,}
\usepackage{amsmath,amsfonts}
\usepackage{amssymb}
\usepackage{physics,amsmath}
\usepackage{array}
\usepackage{bm}
\usepackage{mathrsfs}
\usepackage{dsfont}
\usepackage{siunitx}
\usepackage{ifthen}
\usepackage{mathtools}
\usepackage{relsize}

\usepackage{amsthm}
\usepackage{csquotes}
\usepackage{caption}
\usepackage{subcaption}
\usepackage{graphicx}
\usepackage{mathtools}
\usepackage{xcolor}
\usepackage{tikz}
\usetikzlibrary{external}
\usetikzlibrary{chains}
\usepackage{makecell}
\usepackage{textcomp}
\usepackage{stfloats}
\usepackage{lipsum}  
\usepackage{bm}
\usepackage{url}
\usepackage{verbatim}
\usepackage{graphicx}
\usepackage{cite}
\usepackage{pdfpages}
\usepackage{relsize}
\usepackage{xcolor}
\usepackage[hidelinks]{hyperref} 

\hypersetup{
    colorlinks=false, 
    linkbordercolor={1 0 0}, 
    citebordercolor={0 1 0}, 
    pdfborder={0 0 1} 
}

\usepackage{tikz}
\usetikzlibrary{3d}
\usetikzlibrary{arrows.meta}

\newcommand{\vect}[1]{\mathbf{#1}}

\usepackage{courier} 
\usepackage{algorithm} 
\usepackage{algpseudocode} 
\usepackage{comment}

\begin{document}

\title{Neural-Inspired Multi-Agent Molecular Communication Networks for Collective Intelligence}

\author{\IEEEauthorblockN{Boran A. Kilic\IEEEauthorrefmark{1} \IEEEauthorrefmark{3} and Ozgur B. Akan\IEEEauthorrefmark{2}\IEEEauthorrefmark{3}}

\IEEEauthorblockA{\IEEEauthorrefmark{1} Electrical and Electronics Engineering Department, Bogazici University, Istanbul, Turkey}

\IEEEauthorblockA{\IEEEauthorrefmark{2}Center for Next-generation Communications (CXC), Koç University, Istanbul, Turkey}

\IEEEauthorblockA{\IEEEauthorrefmark{3}Internet of Everything (IoE) Group, Electrical Engineering Division, University of Cambridge, UK}

\IEEEauthorblockA{E-mail: boran.kilic@std.bogazici.edu.tr, oba21@cam.ac.uk}}


\maketitle

\begin{abstract}
Molecular Communication (MC) is a pivotal enabler for the Internet of Bio-Nano Things (IoBNT). However, current research often relies on “super-capable” individual agents with complex transceiver architectures that defy the energy and processing constraints of realistic nanomachines. This paper proposes a paradigm shift towards “collective intelligence,” inspired by the cortical networks of the biological brain. We introduce a decentralized network architecture where simple nanomachines interact via a diffusive medium using a threshold-based firing mechanism modeled by Greenberg-Hastings (GH) cellular automata. We derive fixed-point equations for steady-state populations via mean-field analysis and validate them against stochastic simulations. We demonstrate that the network undergoes a second-order phase transition at a specific activation threshold. Crucially, we show that both pairwise and collective mutual information peak exactly at this critical transition point, confirming that the system maximizes information propagation and processing capacity at the “edge of chaos.”
\end{abstract}

\begin{IEEEkeywords}
Molecular Communication, Internet of Bio-Nano Things, Complex systems, criticality.
\end{IEEEkeywords}

\section{Introduction}

\IEEEPARstart{M}{olecular} Communication (MC) has established itself as a promising paradigm for information exchange in nanonetworks, particularly within the Internet of Bio-Nano Things (IoBNT) domain. While the fundamental limits of point-to-point diffusion channels have been extensively studied, the field is currently shifting towards understanding large-scale, collective interactions.

Despite the extensive literature on channel modeling and transceiver architectures, a significant disconnect remains between theoretical models and practical realizability. First, the vast majority of existing works investigate Single-Input Single-Output (SISO) links or simple geometric arrays in the Multiple-Input Multiple-Output (MIMO) case \cite{body_area_nanonets, 7060516,kilic2025}. These configurations imply a structured, deterministic topology that is unrepresentative of real-life biological environments, where nanomachines are often distributed randomly throughout a medium. Second, current theoretical frameworks frequently rely on idealized assumptions regarding the capabilities of individual agents. These include perfect temporal synchronization, high-precision molecule counting, and the hardware capacity to perform complex signal processing algorithms. Current and near-future nanotechnology, however, limits us to the production of simple, resource-constrained agents incapable of such complex individual tasks.

To bridge this gap, we propose a paradigm shift from "super-capable individual agents" to a "network of simple, numerous agents." This approach draws direct inspiration from the most robust biological network known: the cortical networks of the brain. In the cortex, complex information processing emerges not from the intelligence of a single neuron, but from the collective dynamics of millions of interconnected, simple units.

\begin{figure}[htbp]
    \centering
    \includegraphics[width=\linewidth]{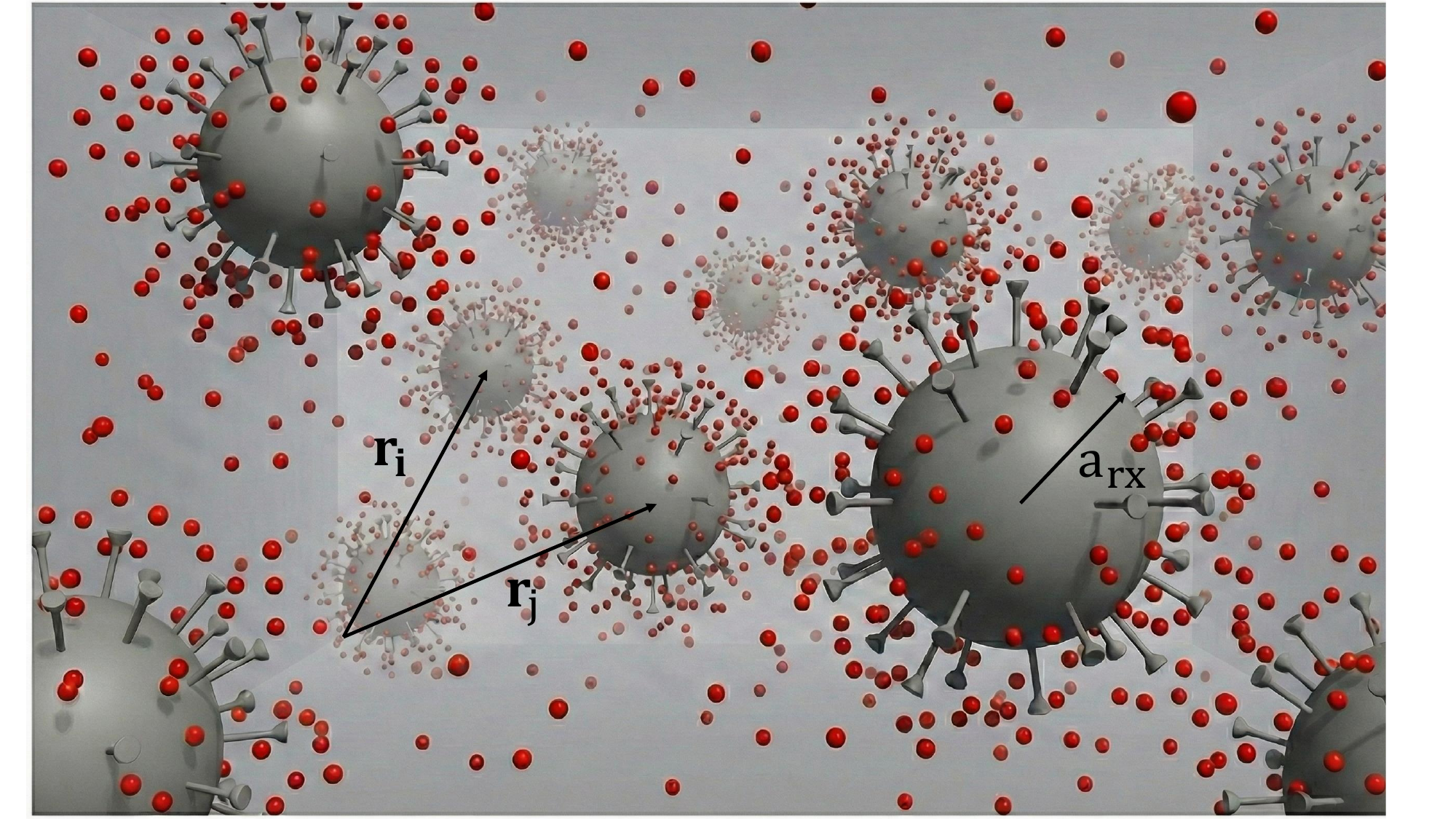}
    \caption{Proposed system model contrasting traditional point-to-point MC with the proposed bio-inspired network of excitable agents.}
    \label{fig:system_sketch}
\end{figure}

In this work, we introduce a decentralized architecture where agents act as transceivers analogous to neurons. Unlike the isolated links found in traditional MC, our agents are distributed randomly and operate on a threshold-based firing mechanism: if the concentration of molecules absorbed during a transmission interval exceeds a threshold, the agent ``fires'' (releases molecules) in the subsequent interval. This mechanism allows us to leverage the nonlinear nature of molecular diffusion, a property recently identified as highly beneficial for reservoir computing\cite{Uzun2025Molecular}. Crucially, reservoir computing paradigms rely on the system operating within a critical regime or at the ``edge of chaos,'' a property we demonstrate is achievable in this diffusive network. While recent studies in organoid intelligence \cite{Cai2023Brain} and biological reservoir computing \cite{Cucchi2021Reservoir} utilize biological tissue for this purpose, no prior work has fully incorporated this collective computing paradigm into  networks of molecular agents communicating via diffusion.

To mathematically formalize these dynamics, one must select an appropriate neuron dynamics model. The computational neuroscience literature offers various continuous-variable models. The Integrate-and-Fire (IF) models and their variants (exponential or quadratic IF) describe neurons that integrate synaptic input until a threshold is reached \cite{Cessac2008, FourcaudTrocm2003, Trappenberg2022}. Alternatively, Spike-Response Models (SRM) utilize kernels to determine spiking probability \cite{SRM}. However, these models rely on continuous membrane-potential dynamics and require costly numerical integration, making them less suitable for analytically modeling the collective interactions of massive, discrete molecular swarms.

Consequently, we adopt an excitable cellular-automaton approach: the Greenberg-Hastings (GH) model \cite{GreenbergHastings1964}. The GH model abstracts the neuron into discrete states—quiescent, excited, and refractory—updating synchronously based on local interactions. This discrete formalism naturally captures the essential wave propagation and refractory behaviors observed in cortical tissue \cite{Durrett1993} while remaining mathematically tractable for molecular diffusion scenarios.

\section{Methods}
In this section, we detail the proposed diffusive network and introduce key system parameters. We also derive the state update equations governing network dynamics.

Each spherical transceiver unit is modeled as a network agent: a communication node capable of receiving and transmitting information molecules based on local decision rules. The network consists of $M$ agents spatially distributed according to a specified geometry. We assume their relative positions remain constant during the interval of interest, with the $i$-th agent located at $\vect{r_i}$. 

We assume molecular motion is governed by free diffusion in a 3D unbounded medium with an isotropic, homogeneous diffusion coefficient $D$. Furthermore, we assume all agents are time-synchronized. The information transmission process is discretized into signaling intervals of duration $T_s$. All molecule counting and emission events occur at discrete time steps $kT_s$ within a duration negligible compared to $T_s$, while free diffusion occurs in the interim, consistent with standard Molecular Communication (MC) setups.

In contrast to Single-Transmitter Single-Receiver (SISO) channel models, selecting a receiver model for a transceiver ensemble presents unique challenges. We employ fully absorbing receivers, as they are physically plausible; absorption occurs on the surface, leaving the internal spherical volume available for molecule storage, computing, and energy harvesting \cite{transmitter_and_receiver}, unlike passive receiver model. We assume the agent distribution is sufficiently sparse such that no agent significantly obstructs molecule transmission between any pair of agents.

\subsection{Arrival Modeling}
The Channel Impulse Response (CIR) between the $i$-th and $j$-th transceiver for a fully absorbing receiver is given by
\begin{equation}
\begin{aligned}
h^{(F)}_{ij}(t)=\frac{a_{\mathrm{rx}}\left(|\mathbf{r}_i - \mathbf{r}_j|-a_{\mathrm{rx}}\right)}{t |\mathbf{r}_i - \mathbf{r}_j| \sqrt{4 \pi D t}} \exp \left(-\frac{\left(|\mathbf{r}_i - \mathbf{r}_j|-a_{\mathrm{rx}}\right)^2}{4 D t}\right) .
\end{aligned}
\end{equation}
The cumulative probability that a molecule will be absorbed by time $t$, then, is calculated as \cite{transmitter_and_receiver}
\begin{equation}
    g_{ij}(t)=\int_{t^{\prime}=0}^t h^{(F)}_{ij}\left(t^{\prime}\right) \mathrm{d} t^{\prime}=\frac{a_{\mathrm{rx}}}{|\mathbf{r}_i - \mathbf{r}_j|} \operatorname{erfc}\left(\frac{|\mathbf{r}_i - \mathbf{r}_j|-a_{\mathrm{rx}}}{\sqrt{4 D t}}\right)
\end{equation}
where $a_{rx}$ is the radius of the receiver, $D$ is the diffusion coefficient and $|\mathbf{r}_i-\mathbf{r}_j|$ is the distance between $i$-th and $j$-th transceiver. It follows that a molecule emitted $k$ intervals before will be absorbed by the fully absorbing receiver with probability $P_{ij}[k]=g_{ij}(kT_s)-g_{ij}\left((k-1)T_s \right)$.

Although the number of detected molecules is binomially distributed, analytical solutions for binomial random variables are often intractable. Thus, it is common practice to model this distribution as Gaussian, which yields negligible discrepancy in the regime of large molecule counts. Accordingly, the number of molecules emitted by the $j$-th agent and received by the $i$-th agent
\begin{equation}\label{binomial_to_gaussian_arrival}
    N^{(R)}_{ij}[k] \sim \mathcal{N}(\mu_{ij}[k], {\sigma_{ij}}^2[k])  \ ,
\end{equation}
with
\begin{equation}\label{eq:expected_num_of_mols}
    \mu_{ij}[k] = N_0 \sum_{\ell =1}^{L} P_{ij}[\ell]\ x_j[k - \ell ] ,
\end{equation}
\begin{equation}\label{eq:standart_dev_of_mols}
   {\sigma_{ij}}^2[k] = N_0 \sum_{\ell=1}^{L} P_{ij}[\ell]\ (1 - P_{ij}[\ell])\ x_j[k - \ell ] ,
\end{equation}
where $N_0$ is the fixed number of molecules released, $L$ is the channel memory, and $x_j[k] \in \{0,1\}$ is the state of $j$-th agent representing whether the agent is excited or not, at the beginning of $k$-th signalling interval. 

Formally, the number of molecules absorbed depends on the history of entire past $L$ states $x_i[k-L:k-1]$. Moreover, each agent will receive some molecules from every other agent. Thus, the total number of molecules absorbed conditioned on the past $L$ states of all agents in the network is
\begin{equation}
    \begin{aligned}
    & N^{(R)}_{i}[k] \ | \ \mathbf{x}[k-L:k-1] \sim \\& \mathcal{N} \Bigl(\mu_{i}[k] \ | \ \mathbf{x}[k-L:k-1], {\sigma_{ij}}^2[k] \ | \ \mathbf{x}[k-L:k-1] \Bigr),
    \end{aligned}
\end{equation}
with
\begin{equation}\label{eq:expected_num_of_mols_cond}
    \begin{aligned}
    \mu_{i}[k] \ | \ \mathbf{x}[k-L:k-1] &= N_0 \sum_{\ell =1}^{L} \sum_{j=1}^{M} P_{ij}[\ell]\ x_j[k - \ell ], \\ 
    \boldsymbol{\mu}[k] \ | \ \mathbf{x}[k-L:k-1] &= N_0 \sum_{\ell =1}^{L} \mathbf{P}[\ell] \mathbf{x}[k-\ell],
    \end{aligned}
\end{equation}
\begin{equation}\label{eq:standart_dev_of_mols_cond}
\begin{aligned}
   {\sigma_{i}}^2[k] \ &| \ \mathbf{x}[k-L:k-1]\\ &= N_0 \sum_{\ell=1}^{L} \sum_{j=1}^{M} P_{ij}[\ell]\ (1 - P_{ij}[\ell])\ x_j[k - \ell ], \\
   \boldsymbol{\sigma}^2[k] \ &| \ \mathbf{x}[k-L:k-1] \\ &=N_0  \sum_{\ell=1}^{L} \Bigl(\mathbf{P}[\ell] \odot (\mathds{1}-\mathbf{P}[\ell] \Bigl) \mathbf{x}[k-\ell],
\end{aligned}
\end{equation}
where $\mathbf{P}[k]$ is the symmetric connectivity matrix representing the response function between agents,  $\mathbf{x}[k] = [x_1[k] \ x_2[k] \ldots x_M[k]]^T $, $\mathds{1}$ is the all-ones matrix, and $\odot$ represents element-wise multiplication. Diagonal elements of $\mathbf{P}[k]$, representing self-interactions, are set to zero to prevent a self-amplified, unbounded response.

\subsection{Greenberg-Hastings Model}
According to the GH model \cite{greenberg_hastings_1978, GH_in_neurons}, an agent (or node) exists in 3 states: refractory $(R)$, excited $(E)$ and quiescent ($Q)$. Transition rules between states are defined as 

\begin{equation}\label{eq:update_rules}
\begin{aligned}
    R \to Q & \quad \text{with probability } \tilde{r}_2 \\
    E \to R & \quad \text{with probability } 1 \\
    Q \to E & \quad \text{if} \ N^{(R)}>T \text{ or with probability } \tilde{r}_1 \\
\end{aligned}
\end{equation}
where $N^{(R)}$ is the total number of molecules received and $T$ is the activation threshold, a critical parameter controlling network dynamics. Upon excitation, the agent isotropically releases $N_0$ molecules into the medium. The non-linearity inherent in this threshold-dependent excitation mechanism, which is analogous to neuronal dynamics, is a key feature of the network.

Let $\tilde{x}_i[k]$ be the state of $i$-th agent at time step $k$ where
\begin{equation}
   \tilde{x}_i[k]= \begin{cases}
            0 \ &,  \ \text{refractory} \\
            1 \ &,  \ \text{excited} \\
            2 \ &,  \ \text{quiescent} \\
            \end{cases} \quad .
\end{equation}
We define the probability of any node being in one of the 3 GH states as:
\[
\begin{aligned}
    r_i[k] &= \text{Pr}\left\{\tilde{x}_i[k]=0\right\} \\
    e_i[k] &= \text{Pr}\left \{\tilde{x}_i[k]=1 \right\} \\
    q_i[k] &= \text{Pr}\left \{\tilde{x}_i[k]=2 \right\}  \quad .\\ 
\end{aligned}
\]
State evolution is generally stochastic due to spontaneous activation probability $\tilde{r}_2$, external excitation probability $\tilde{r}_1$ and random initialization of neurons. However, independent of whether the simulations are stochastic or deterministic, we consider $\tilde{x}_i[k]$ as a random variable in the state space of all possible configurations. Notice that $\tilde{x}_i[k]=0$ and $\tilde{x}_i[k]=2$ map to $x_i[k]=0$ and $\tilde{x}_i[k]=1$ maps to $x_i[k]=1$.

Using the transition dynamic of GH model \ref{eq:update_rules}
\begin{align}\label{eq:e_to_r}
    r_i[k+1] &= e_i[k]+(1- \tilde{r}_2) \cdot r_i[k]  \\ 
    q_i[k+1] &= \tilde{r}_2 \cdot r_i[k] + \left(1-\gamma_i[k] \right) \cdot q_i[k]  \\
    e_i[k+1]   &= \left(\tilde{r}_1+\gamma_i[k] \right) \cdot q_i[k] \label{eq:q_to_ex}
\end{align}
where
\begin{equation}\label{eq:spont_excitation_prob}
    \gamma_i[k] = \text{Pr} \left\{N^{(R)}_i[k] > T \right\} \ = 1- \Phi\left(\frac{T-\mu_i[k]}{\sigma_i[k]}\right)
\end{equation}
where $\Phi(\cdot)$ is the normal cumulative distribution function (CDF).

Noting that $\mathbf{e}[k] = \mathbb{E} \Bigl[\mathbf{x}[k]\Bigr]$ averaged over all possible configurations, unconditional mean of Gaussian arrivals in \ref{eq:expected_num_of_mols_cond} becomes
\begin{equation}\label{eq:mu_i}
    \mathbb{E} \Bigl[ N^{(R)}_{i}[k] \Bigl]= \mu_i[k] =N_0 \sum_{\ell =1}^{L} \mathbf{P}_i[\ell] \cdot\mathbf{e}[k-\ell]
\end{equation}
where $\mathbf{P}_i[\ell]$ is the $i$-th row of $\mathbf{P}[\ell]$.
The nconditional variance, by \textit{the law of total variance}, becomes
\begin{equation}
\begin{aligned}
    \text{Var}[N_i^{(R)}[k]] = \sigma_i^2[k]&= \underbrace{\mathbb{E}\biggl[\text{Var}\Bigl(N^{(R)}_{i}[k] \ | \ \mathbf{x}[k-L:k-1]\Bigr) \biggr]}_{\Sigma_i^{(1)}[k]} \\   &+ \underbrace{\text{Var}\biggl[\mathbb{E}\Bigl(N^{(R)}_{i}[k] \ | \ \mathbf{x}[k-L:k-1]\Bigr)\biggr]}_{\Sigma_i^{(2)}[k]} 
\end{aligned}
\end{equation}
where the first term $(\Sigma_i^{(1)}[k])$ represents the variance of Gaussian arrivals from molecular diffusion, and the second term $(\Sigma_i^{(2)}[k])$ represents the variance due to the Greenberg-Hastings process. The first term is obtained by taking the expected value of eq. \ref{eq:standart_dev_of_mols_cond} as
\begin{equation}\label{eq:sigma_1}
    \Sigma_i^{(1)}[k] = \sum_{j,l} N_0  \ e_j[k-\ell] P_{ij} [\ell]\Bigl(1-P_{ij}[\ell] \Bigr) .
\end{equation}
The second term is obtained by calculating the variance of eq. \ref{eq:expected_num_of_mols_cond} 
\begin{equation}\label{eq:sigma_2_cov}
    \begin{aligned}    
    &\Sigma_i^{(2)}[k] = \text{Var} \Bigr[ \sum_{j,l} N_0 P_{ij}[\ell] x_j[k-\ell] \Bigr] \\
    &= \sum_{j,l}N_0^2P_{ij}^2[\ell] \ \text{Var}[x_j[k-\ell]] \\ &+2 \sum_{(j,\ell)<(j',\ell')} N_0^2 P_{ij}[\ell]P_{ij'}[\ell'] \text{Cov}\Bigl[x_j[k-\ell],x_{j'}[k-\ell']\Bigr] 
    \end{aligned}
\end{equation}
where $(j,\ell)<(j',\ell')$ represents the sum over all ordered pairs $(j,\ell)$ when either $j=j'$ and $\ell<\ell'$ or $\ell=\ell'$ and $j<j'$. 

Since $x_j[k-\ell]$ are independent across nodes, and assuming temporal dependence is negligible compared to other terms, the covariance term is neglected for analytic tractability. This assumption is validated in the Results section via simulations of the GH model. Finally, noting that $x_j[k] \sim \text{Bernoulli}(e_j[k])$, the total variance is derived as
\begin{equation}\label{eq:var_i}
\begin{aligned}
   \text{Var}[N_i^{(R)}[k]] & \approx \sum_{j,\ell} N_0e_j[k-\ell]P_{ij}[\ell]  (1-P_{ij}[\ell]) \\ & +\sum_{j,\ell} N_0^2 P_{ij}^2[\ell]e_j[k-\ell]  (1-e_j[k-\ell])  
   \end{aligned}
\end{equation}

\subsection{Steady-State Analysis}

As $t\rightarrow \infty$ or practically for a large enough time, state probabilities will settle to their steady-state values. 
\begin{equation}
    \{e, r, q\}_i[k] = \{e, r, q\}_i[k+1] = \{e^*, r^*, q^*\}_i \quad \text{as} \rightarrow \infty
\end{equation}
and also $\gamma_i[k] \rightarrow \gamma_i^*$. Then equations \ref{eq:e_to_r} become:
\begin{equation}  
    r_i^*  = e_i^*+(1-r_2) \cdot r_i^*  \Rightarrow r_i^* = \frac{e_i^*}{r_2}    
\end{equation}
Using $q_i^* + e_i^* + r_i^*=1$, we get
\begin{equation}\label{eq:q_i_star}
    q_i^* = 1 - e_i^*- r_i^* = 1-e_i^* \left(1+\frac{1}{\tilde{r_2}}\right) .
\end{equation}
Finally, rewriting \ref{eq:q_to_ex}
\begin{equation} \label{eq:e_i*}
    \begin{aligned}
    e_i^* &= (\tilde{r}_1 + \gamma_i^*) \cdot \left( 1-e_i^* \left(1+\frac{1}{\tilde{r_2}}\right) \right) \\ &\Rightarrow e_i^* = \frac{\tilde{r}_1 + \gamma_i^*}{1+ (\tilde{r}_1 + \gamma_i^*)\left(1+1/\tilde{r_2}\right)} \quad .
    \end{aligned}
\end{equation}
Generalizing to all $M$ agents, steady state probabilities can be found from the $M$-dimensional fixed-point problem
\begin{equation}\label{fixed_point}
   \mathbf{e}^* = \frac{\tilde{r}_1 + \Gamma(\mathbf{e}^*)}{1+ (\tilde{r}_1 + \Gamma(\mathbf{e}^*))\left(1+1/\tilde{r_2}\right)} \quad .
\end{equation}
where $\Gamma(\mathbf{e}^*)$ is the steady-state excitation vector directly derived from eq. \ref{eq:spont_excitation_prob}. Accumulating all past $L$ contributions from each state with probability $\mathbf{e}^*$,
\begin{equation}\label{eq:gamma_e_star}
    \Gamma(\mathbf{e}^*)=  1- \Phi\left(\frac{T-\boldsymbol{\mu}^*}{\boldsymbol{\sigma}^*}\right)
\end{equation}
where
\begin{equation}
    \boldsymbol{\mu}^* = N_0 \left(\sum_{\ell=1}^{L} \mathbf{P}[\ell]\right) \cdot \mathbf{e}^*   
\end{equation}
\begin{align}
    {\boldsymbol{\sigma}^*}^2 &= N_0 \Bigl( \sum_{\ell=1}^{L} \mathbf{P}[\ell] \odot (\mathds{1}-\mathbf{P}[\ell]) \Bigl) \cdot \ \mathbf{e}^* \\
    & + N_0^2 \Bigl( \sum_{\ell=1}^{L} \mathbf{P}^2[\ell] \Bigr) \cdot \mathbf{e}^*\cdot (1-\mathbf{e}^*)
\end{align}

\section{Results}
To analyze how the network evolves, we simulate the dynamics using MATLAB. In the simulations, we distribute $N=100$ agents homogeneously according to a Poisson point process within a sphere of $R=20 \ \mathrm{\mu m}$. We calculate the pairwise distances between nodes to pre-calculate the connectivity matrix $\mathbf{P}_{hit}$. At each time step, the received molecule count for every node is sampled from a Gaussian distribution, with the mean and variance derived from $\mathbf{P}_{hit}$ and the past $L$ states. Subsequently, GH transition rules are applied, and molecule counts for nodes entering or remaining in the refractory state ($R$) are reset to zero.

\begin{figure}[htbp]
    \centering
    \includegraphics[width=\linewidth, trim={3cm 0 1cm 0}, clip]{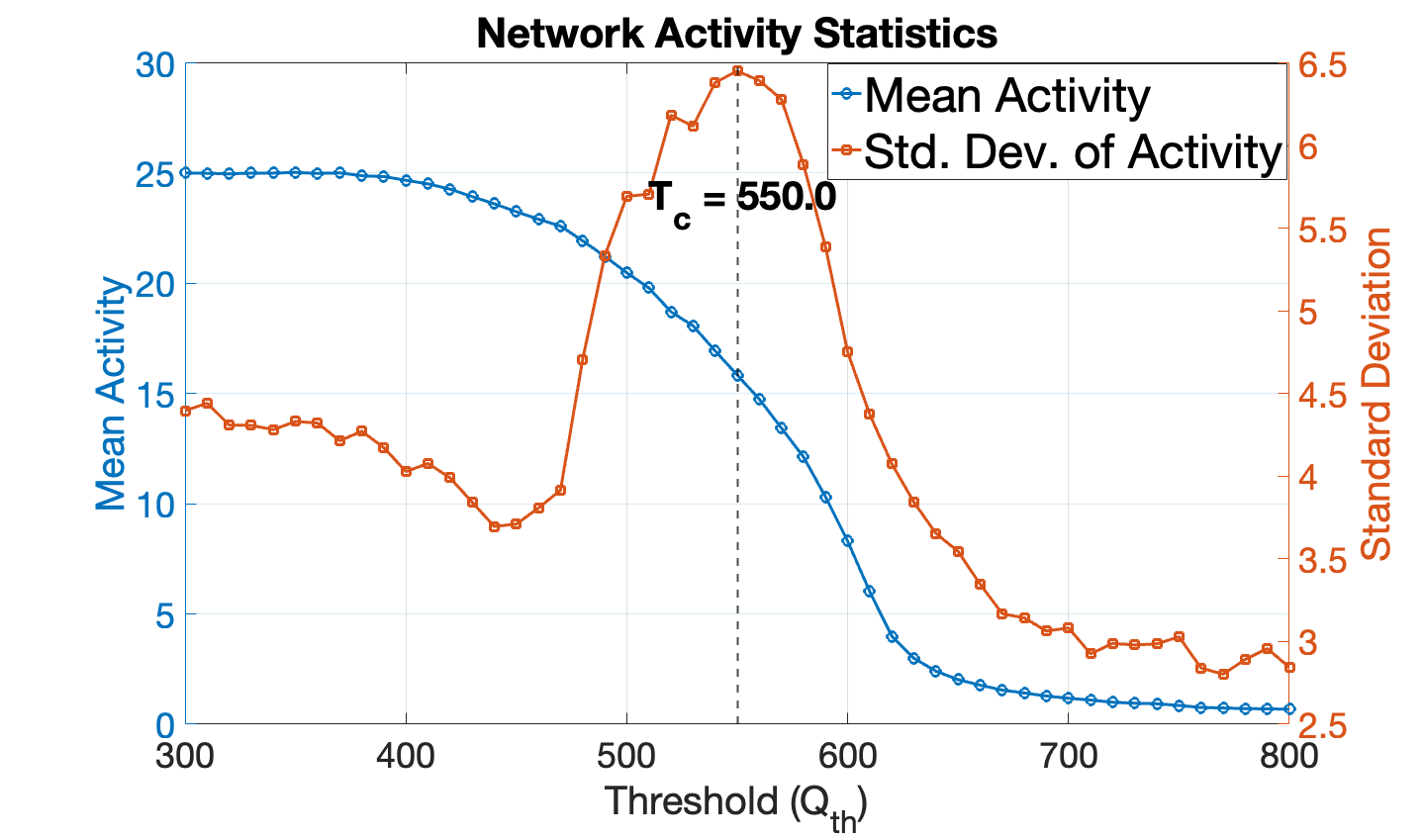}
    \caption{Average mean and standard deviation of activity in the system with $T_s=1 \ \mathrm{s}$, $D=79.4 \ \mathrm{\mu m} $, $Q_0=100$, $r=4$, $L=5$, $\tilde{r}_1=0$, $\tilde{r}_2=0.5$}
    \label{fig:mean_std_activity}
\end{figure}

In Fig. \ref{fig:mean_std_activity}, we show the mean and standard deviation of excited activity. As the threshold $Q_{th}$ increases, mean activity decreases. Around $Q_{th} \sim 500$, the network undergoes a second-order phase transition, where mean activity rapidly drops to 0. The standard deviation of activity serves as the main indicator of network susceptibility, measuring how activity gives rise to large fluctuations in the network \cite{Timme2016}. The standard deviation peaks in the middle of this phase transition, indicating a critical threshold for the network. The results categorize the network into three distinct regimes: a sub-critical 'saturation' phase ($Q_{th} < 450$) where noise dominates, a super-critical 'silence' phase ($Q_{th} > 600$) where activity dies out, and the critical phase ($Q_{th} \approx 500$) where the system achieves optimal computational properties.

\begin{figure}[htbp]
    \centering
    \includegraphics[width=\linewidth, trim={1cm 0  1cm 0}, clip]{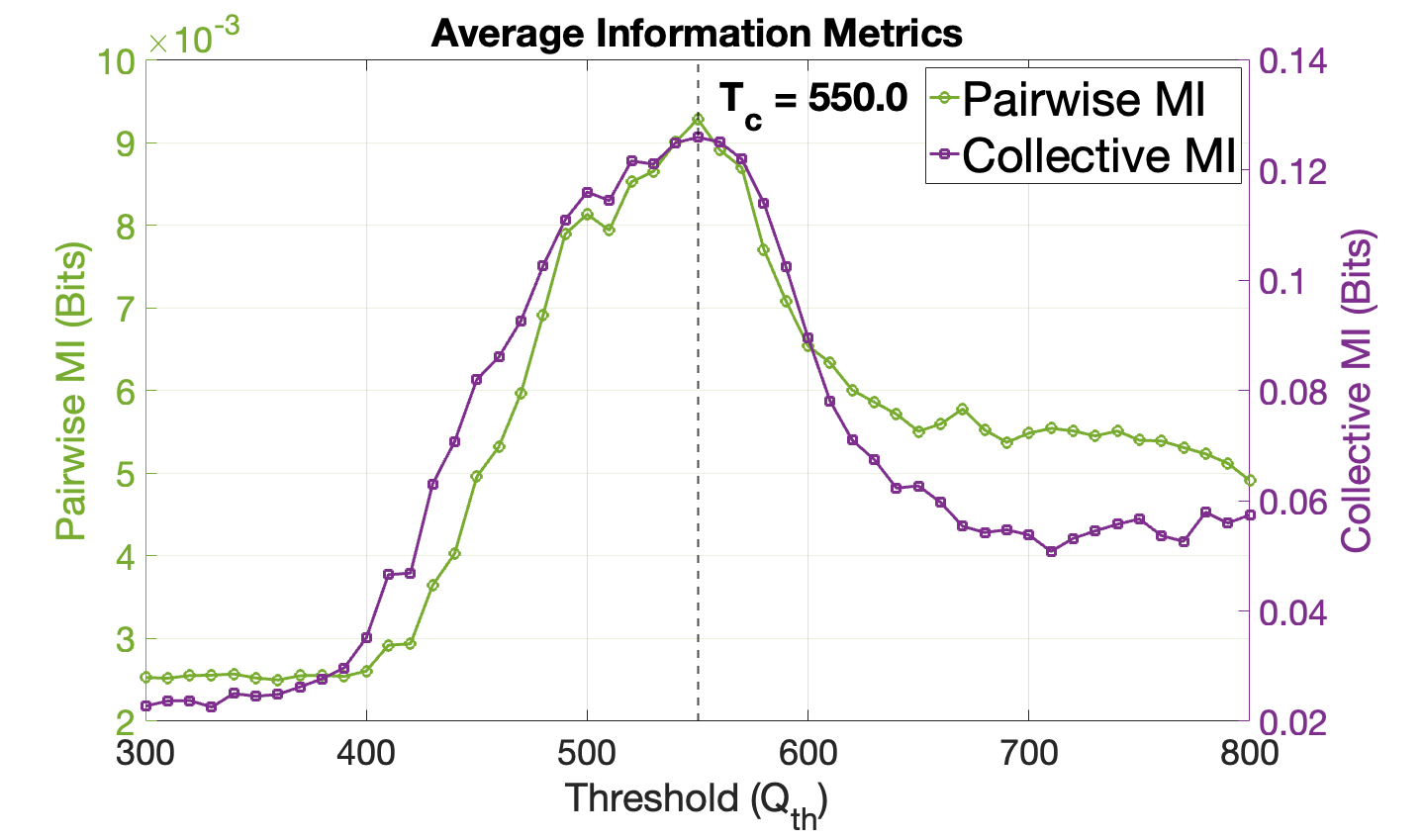}
    \caption{Pairwise and collective mutual information in the network. Same simulation parameters as Fig. \ref{fig:mean_std_activity} are used.}
    \label{fig:pairwise_collective_mi}
\end{figure}

In Fig. \ref{fig:pairwise_collective_mi}, we evaluate the pairwise and collective mutual information (MI) of the system, defined as:
\begin{equation}
    \bar{I}_{\text{pair}} = \binom{N}{2}^{-1} \sum_{i < j} I(X_i; X_j), \quad \bar{I}_{\text{coll}} = \frac{1}{N} \sum_{i=1}^{N} I(X_i; M_i)
\end{equation}
where $M_i$ represents the aggregate activity of all agents excluding the $i$-th agent. Metrics are computed using the excited state history averaged over $n=10$ simulation runs. We observe that both MI metrics peak at the same threshold identified in the activity analysis, corroborating the location of the critical point \cite{Beggs_2022}. This maximization of information exchange serves as a quantitative signature of collective intelligence, where the network as a whole processes information more effectively than the sum of its parts.

For the mean-field analysis, we solve the fixed-point equation \ref{fixed_point} using the fixed-point iteration method with under-relaxation:
\begin{equation}
    \mathbf{e}^*_{k+1} = \lambda \cdot g( \mathbf{e}^*_{k})+ (1-\lambda)  \mathbf{e}^*_{k}
\end{equation}
where $g(\cdot)$ is the right-hand side of Eq. \ref{fixed_point} and $\lambda=0.9$ with tolerance $\epsilon=10^{-12}$.

\begin{figure}[htbp]
    \centering
    \includegraphics[width=\linewidth, trim= {1cm 0 3cm 0}, clip]{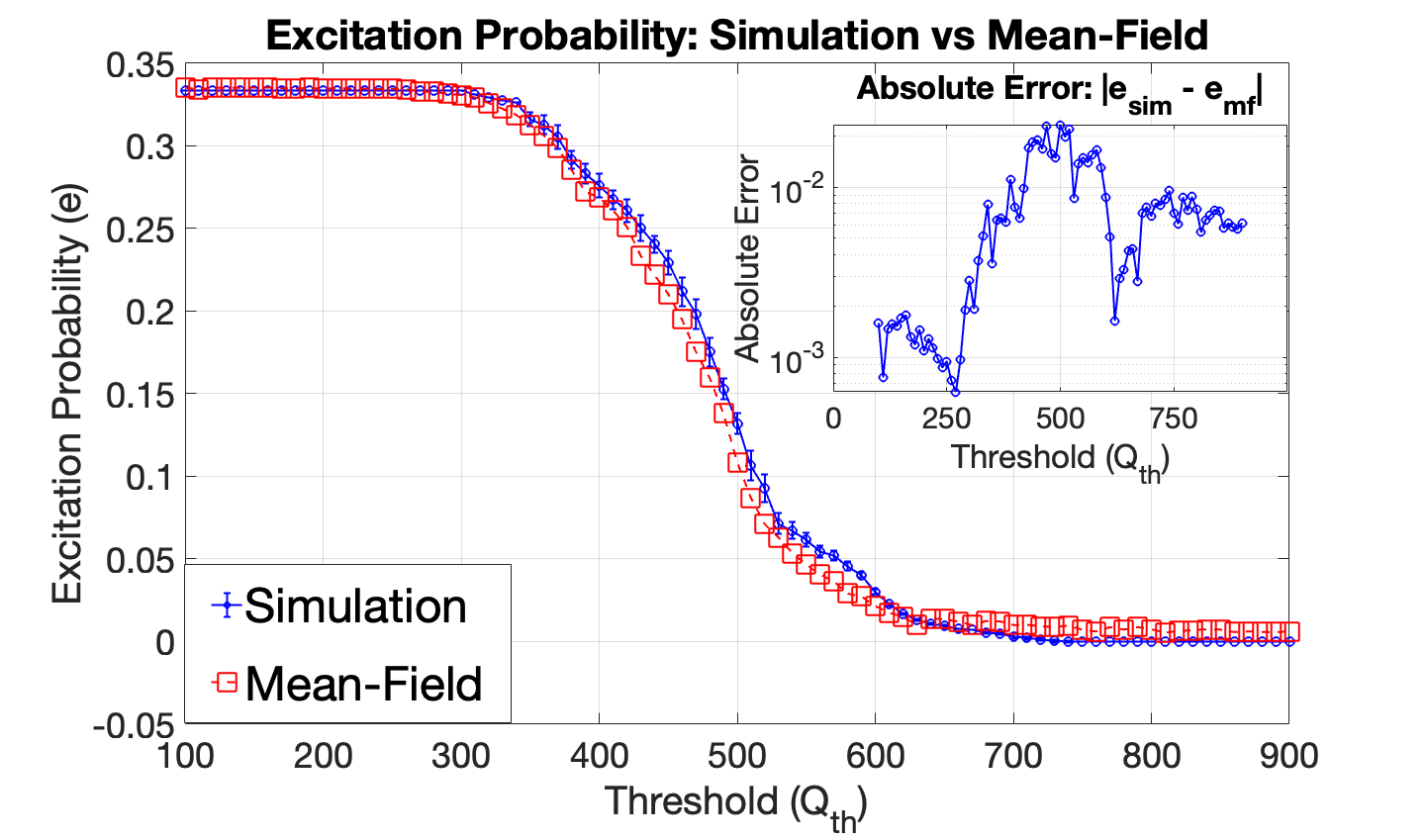}
    \caption{Theoretical and simulated excitation probabilities. Simulation parameters are same as Fig. \ref{fig:mean_std_activity} except $r_2=1$}.
    \label{fig:mean_field_comp}
\end{figure}

To validate the steady-state analysis, we perform $100$ simulation runs, each lasting $500$ time steps, for every $Q_{th}$ value. The mean excitation probability is calculated as the time-averaged excited population, averaged across all runs. Fig. \ref{fig:mean_field_comp} compares the simulated and theoretical mean excitation probabilities over a range of thresholds. In general, the theoretical estimate proves accurate, despite the instability caused by stochastic arrivals and the randomness of the GH process. However, the divergence between simulation and theory in Fig. \ref{fig:mean_field_comp} is not random; it aligns precisely with the transition region. This confirms that the mean-field assumption of node independence breaks down exactly when the system maximizes information transfer, as criticality is inherently defined by strong, long-range correlations between agents. Consequently, the covariance term in Eq. \ref{eq:sigma_2_cov}—neglected in the mean-field approximation—becomes non-negligible, causing the observed discrepancy. Nevertheless, the proposed analysis remains valid for approximating the macroscopic behavior of the network.

In this regime, signal propagation neither dies out quickly nor amplifies chaotically but persists in a stable, self-sustaining state at the ``edge of chaos.'' This criticality maximizes the network's dynamic range, offering a robust framework for harnessing collective intelligence for both information transmission and distributed molecular computing.

\section{Conclusion}
This paper proposes a shift toward "collective intelligence" in the IoBNT by utilizing simple, decentralized nanomachines instead of complex individual agents. By modeling these interactions through the GH cellular automata, we demonstrate that tuning activation thresholds allows the network to reach a critical regime that maximizes information propagation.

Unlike standard graph-based networks where edge weights are static, the communication links here are inherently stochastic, meaning even identical initial network configurations can lead to diverging dynamic behaviors. Furthermore, the diffusive channel introduces significant memory—molecules linger in the medium, creating Inter-Symbol Interference (ISI) that prevents the system from being analyzed as a simple Markovian process. While these aspects complicate the mathematical analysis, they also endow the system with computational power akin to biological neural networks. The residual molecular concentration mimics synaptic facilitation and depression, where past activity modulates future response , while the channel noise can be leveraged via stochastic resonance, a phenomenon where an optimal level of noise actually enhances the detection of weak signals and maximizes information transmission. We rely on the principle that, despite local fluctuations, the network's collective statistics will converge over many iterations to a stable fixed point, achieving optimal information processing at the edge of chaos if the parameters are set correctly.

Future efforts are planned to explore different network geometries, perturbation analysis, and provide further theoretical and simulation-based proof of the system's operation within critical regimes.

\bibliographystyle{ieeetr}
\bibliography{references}

\end{document}